\documentclass[aps,prl,twocolumn,10pt,altaffilletter,nolongbibliography,numerical,flushbottom,secnumarabic,superscriptaddress,floatfix,nobibnotes]{revtex4-2}

\usepackage{graphicx}
\usepackage{braket}
\usepackage{amsmath, amssymb}
\usepackage{xfrac}
\usepackage{booktabs}  
\usepackage{comment}
\usepackage[normalem]{ulem}
\usepackage{stix}
\usepackage[breaklinks, pdftex, hyperfootnotes=true, pdfpagelabels, bookmarks, pageanchor]{hyperref}
\usepackage{comment} 

\usepackage{xcolor}

\hypersetup{%
	colorlinks=true, linktocpage=true, pdfstartpage=1, pdfstartview=FitH, pdfborder={0 0 0},%
	breaklinks=true, pdfpagemode=UseNone, pageanchor=true, pdfpagemode=UseOutlines,%
	plainpages=false, bookmarksnumbered, bookmarksopen=true, bookmarksopenlevel=1,%
	hypertexnames=true, pdfhighlight=/O,
	urlcolor=blue, linkcolor=blue, citecolor=blue,
	}
\usepackage{orcidlink}

\newcommand\identity{1\kern-0.25em\text{l}}

\graphicspath{{figs/}}
 
\usepackage{soul}
\newcommand{\psc}[1]{\textcolor{blue}{#1}}

\begin{document}

\title{Geometric heat pumping on a quantum processor}

\author{Ger\'onimo J. Caselli~\orcidlink{0009-0008-2276-3785}}
\affiliation{Facultad de Ciencias Exactas, Ingenier\'ia y Agrimensura, Universidad Nacional de Rosario and Instituto de F\'isica Rosario (CONICET), Rosario, Argentina}
\author{Luis O. Manuel~\orcidlink{0000-0001-7793-1709}}
\affiliation{Facultad de Ciencias Exactas, Ingenier\'ia y Agrimensura, Universidad Nacional de Rosario and Instituto de F\'isica Rosario (CONICET), Rosario, Argentina}
\author{Liliana Arrachea~\orcidlink{0000-0002-7223-4610}}
\affiliation{Centro At\'omico Bariloche, Instituto de Nanociencia y Nanotecnolog\'ia CONICET-CNEA and Instituto Balseiro (8400), San Carlos de Bariloche, Argentina}
\author{Pablo S. Cornaglia~\orcidlink{0000-0003-4991-6573}}
\affiliation{Centro At\'omico Bariloche, Instituto de Nanociencia y Nanotecnolog\'ia CONICET-CNEA and Instituto Balseiro (8400), San Carlos de Bariloche, Argentina}

\begin{abstract}
Geometric heat pumping is the transfer of energy between reservoirs at equal temperature where the heat exchanged in an adiabatic cycle depends only on the driving contour in parameter space. We implement this mechanism on a superconducting quantum processor using a collision model, where a driven qubit interacts repeatedly with two ancilla qubits that are prepared in thermal states and emulate the reservoirs. Energy measurements before and after each interaction provide a calorimetric record of the heat exchanged with each reservoir. By reversing the driving contour, we extract a heat contribution associated with pumping. Its dependence on the period and on the contour follows exact simulations of the collision model, without adjustable parameters, and approaches in the slow-driving regime the geometric limit of the corresponding Lindblad dynamics.
\end{abstract}

\maketitle

{\em Introduction. }
Two reservoirs at the same temperature exchange no net heat. Yet a system coupled in an asymmetric way to both reservoirs and driven around a closed loop in parameter space can
transfer heat from one to the other. This is the thermal counterpart of quantum particle pumping, in which a periodic modulation generates transport without a bias, a mechanism experimentally realized in electronic systems~\cite{Geerligs1990May,Kouwenhoven1991Sep,Pothier1992Jan,Switkes1999Mar,DiCarlo2003Dec,Roche2013Mar,Connolly2013Jun,Buitelaar2008Sep,Jensen1992Nov} and also explored with cold atoms~\cite{Amico2022Nov,Citro2023Feb,Acciai2025Oct}.
In the adiabatic regime, where driving is slow compared with relaxation \cite{Moskalets2004May}, the number of pumped particles per cycle is characterized by a line integral along the contour defined by the periodic driving parameters,  analogous to a Berry phase~\cite{Brouwer1998Oct,Avron2004Aug,Avron2000Oct,Vavilov2001Apr,Victor1984Mar}.
Similar geometric concepts also provide powerful tools for designing and characterizing fast periodic protocols \cite{Takahashi2020Apr,Funo2020Apr}. In systems with nontrivial topology, the pumped charge can furthermore become quantized~\cite{Niu1990Apr}.

Adiabatic quantum heat pumping and its geometric characterization have also been investigated theoretically in driven-qubit systems \cite{Arrachea2023Jan,Bhandari2020Oct,TerrenAlonso2022Feb,Caselli2026May}.
Examples include pumping induced by driven thermal baths \cite{Ren2010Apr}, which is similar to pumping in classical systems \cite{Sinitsyn2007Feb}, as well as power transfer between driving sources \cite{Martin2017Oct,Esin2025Apr}.
When the reservoirs have different temperatures, pumping enables the heat--work conversion underlying continuously operated heat engines and refrigerators~\cite{Ludovico2016Feb,Bhandari2020Oct,TerrenAlonso2022Feb,Caselli2026May}. Unlike discrete-stroke thermal machines~\cite{Rossnagel2016Apr,vonLindenfels2019Aug,Abiuso2020Mar,Cangemi2024Oct}, these devices remain coupled to both reservoirs and must operate against a continuous heat leak, as do thermoelectric machines~\cite{Benenti2017Jun} and adiabatically driven electron motors under voltage bias~\cite{Bustos-Marun2013Aug,Bustos-Marun2019Aug}.

The connection between heat pumping and refrigeration is particularly relevant for quantum
processors, where efficient qubit cooling is essential for accurate
operation. Many proposals use auxiliary qubits as a thermal
machine that cools a target while dissipating heat elsewhere, motivating
bounds on the achievable cooling~\cite{Clivaz2019Oct,BassmanOftelie2024Jul,Xuereb2025Dec}.
Complementary approaches rely on measurements
\cite{Buffoni2019Feb,Buffoni2023Mar,Solfanelli2021Sep}, device architecture
\cite{Portugal2022Nov,Aamir2025Feb,Baruah2025Mar}, or geometric optimization of finite-time dissipation
\cite{Taranto2025Feb}.

Experimental realizations of quantum thermal machines ~\cite{Rossnagel2016Apr,vonLindenfels2019Aug,Aamir2025Feb}
were an important breakthrough in the development of the field of quantum thermodynamics
\cite{Campbell2026Jan}.
They have  established the feasibility of controlling energy changes in few-level systems coupled to effective baths. 
Geometric heat pumping poses a distinct experimental challenge: resolving a contribution governed by the driving contour while distinguishing it from dissipative heating. This requires control over the driving protocol and reservoir interactions, together with measurements of the heat exchanged with each reservoir. Despite recent experimental progress, the measurement of heat remains particularly challenging in quantum devices \cite{Pekola2021Oct}.
Programmable quantum processors provide suitable capabilities
for the implementation of driving protocols, and thermal 
reservoirs by repeated interactions with auxiliary qubits~\cite{Garcia-Perez2020Jan,Cattaneo2021Apr,Erbanni2023Sep} in a quantum collision model~\cite{Ciccarello2022Apr,Cattaneo2022Dec}. Each ancilla is prepared in a thermal state, interacts with the working system, and is then reset before being reused. Measurements before and after the interaction determine its energy change, allowing the ancilla to serve as both a reservoir unit and a calorimeter for the implemented model. 
Repeating this sequence provides a microscopic description of Markovian open-system dynamics and consistent thermodynamic bookkeeping for heat, work, and nonequilibrium steady states
\cite{Karevski2009May,Strasberg2017Apr,Barra2015Oct,Seah2019Oct,Hofer2017Dec,DeChiara2018Nov,Barsky-Giles2026May}. 

In this Letter, we use this approach to implement geometric heat pumping on a superconducting quantum processor. 
A working qubit interacts with two ancillas representing reservoirs at equal temperature. 
Periodic modulation changes both the qubit level splitting and its relative coupling to the reservoirs. 
In the adiabatic regime, the qubit absorbs heat as its splitting decreases and releases heat as it increases; coupling these stages preferentially to different reservoirs produces directed heat transport. 
By reversing the driving contour, we isolate an orientation-odd heat contribution and characterize its dependence on the contour and driving period. Comparison with the discrete collision model and its continuous-time Lindblad limit connects the measured response to geometric pumping, even in the presence of a larger heating background.

{\em Physical model.}
We consider a qubit coupled to two thermal reservoirs, L and R, at the same temperature $T$. Its Hamiltonian is
\begin{equation}
\mathcal H_q(t)=B_x(t)\hat\sigma_q^x+B_z(t)\hat\sigma_q^z,
\end{equation}
where the effective field $\mathbf B(t)=[B_x(t),0,B_z(t)]$ follows a closed path with period $\mathcal T$. 
Its magnitude $B(t)=|\mathbf B(t)|$ determines the energy separation $2B(t)$, while its direction determines the instantaneous energy eigenstates.

Both reservoirs exchange energy with the qubit by inducing transitions between these levels. We choose complementary angular dependences for the transition rates induced by the two reservoirs: at a fixed level splitting, transitions induced by L are strongest when the effective field points along $\hat{z}$ and vanish when it points along $\hat{x}$; the reverse holds for R. Changing the field direction therefore controls the relative strengths of the two thermal contacts. Together with the modulation of the level splitting, this allows the qubit to absorb energy preferentially from one reservoir and release it into the other.
We set $\hbar=k_B=1$ and express energies and times in units of $T$ and $1/T$, respectively.

{\em Collision model.}
 We represent each reservoir by a sequence of independently prepared thermal ancillas \cite{Ciccarello2022Apr}. The drive is divided into $N$ cells of duration $\Delta t=\mathcal{T}/N$, with parameters frozen at the cell midpoints $\overline{t}_j=(j+1/2)\Delta t$, $j=0,\ldots,N-1$.
In cell $j$, we define $B_j=B(\overline{t}_j)$ and  $\hat{\tau}_q^z(\overline{t}_j)=\mathcal{H}_q(\overline{t}_j)/B_j$ as the Pauli operator along the instantaneous effective field. We choose the ancilla quantization axis along the same direction with Hamiltonian ${\cal H}^{(0)}_{\alpha,j}= B_j \hat{\tau}^z_\alpha$.  Each ancilla is prepared in the thermal state
\begin{equation}\label{eq:thermal}
\rho^{\rm th}_{\alpha,j}=|0\rangle\langle 0|p^e_{j}+ |1\rangle\langle 1|(1-p^e_{j}), 
\end{equation} 
where  $p^e_{j}=1/\left[1+e^{2 B_{j}/T}\right]$, and $|0\rangle$ and $|1\rangle$ are the ancilla eigenstateshave with energies $+B_j$ and $-B_j$, respectively. This matching of the energy splitting in the ancilla with the instantaneous one in the driven qubit, allows a collision to exchange an excitation through the partial swap gate
\begin{equation}\label{eq:swap}
  S_{\alpha,j}= \exp\left[ -i \varphi_{\alpha,j}\left(\hat{\tau}^+_\alpha \hat{\tau}^-_q+
\hat{\tau}^-_\alpha \hat{\tau}^+_q\right)\right]
\end{equation}
where the operators $\tau^{\pm}_\alpha,\;\tau^{\pm}_q $ are, respectively, raising/lowering operators in the quantization axis of ${\cal H}^{(0)}_{\alpha,j},\;{\cal H}_q(\overline{t}_j)$.
After each collision the ancilla is reset, so it carries no correlations into the next interaction.

The qubit interacts sequentially with both ancillas on every cell. Each collision  is defined by the interaction with the reservoir acting for over $\Delta t/2$, with the partial swap angle fixed by
\begin{equation}
  \cos^2\varphi_{\alpha,j}= e^{-\Gamma_\alpha(\overline{t}_j)\Delta t/2}.
\end{equation}
The rate scale
\begin{equation}
\Gamma_\alpha(t)=\tfrac{1}{2}\gamma[B(t)] g_\alpha^2 |\boldsymbol{n}_\alpha \times \boldsymbol{n}_B(t)|^2,
\end{equation}
where $\mathbf{n}_B(t)=\mathbf{B}(t)/B(t)=[\sin\vartheta(t),0,\cos\vartheta(t)]$, depends on the strength of the coupling $g_\alpha^2$ and the angular form factor of the reservoir $\alpha$. We consider $\boldsymbol{n}_L= \hat{\boldsymbol{x}}$ and
$\boldsymbol{n}_R= \hat{\boldsymbol{z}}$, in order to represent an asymmetric 
coupling to the baths.
The function $\gamma[B(t)]$ encodes the energy and temperature dependence of the bath-induced transition rates.
We choose $
\gamma[B(t)]= B(t) \coth[B(t)/T]e^{-2B(t)/\omega_c}$, 
which corresponds to Ohmic thermal transition rates with cutoff $\omega_c$ in the continuous time limit derived below. 

{\em Quantum processor implementation.}
\begin{figure}
    \centering
    \includegraphics[width=0.45 \textwidth]{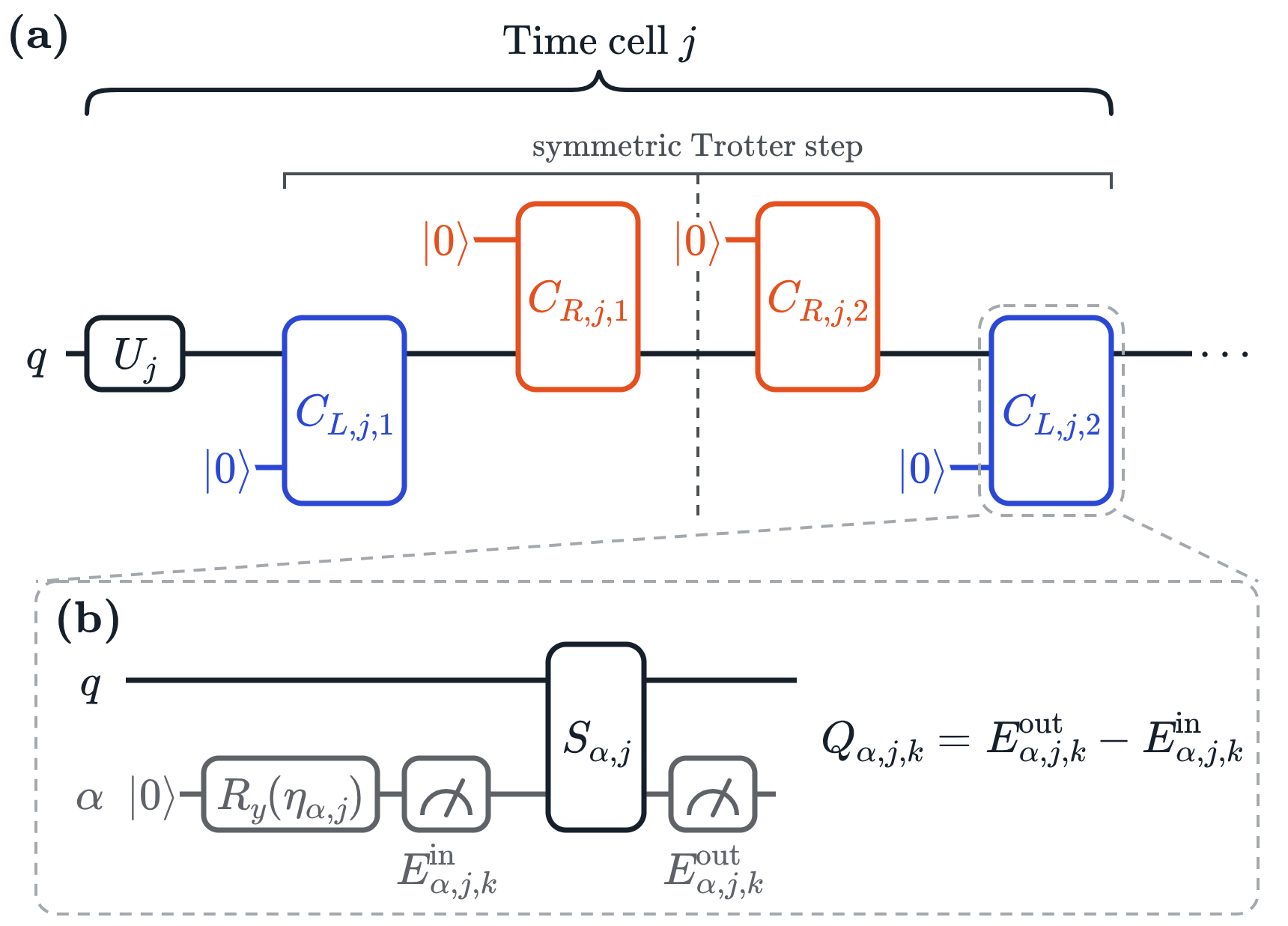}
    \caption{Quantum circuit for one time cell $j$ of the heat pump. 
    (a) The working qubit $q$ evolves under $U_{j}$ and then undergoes the symmetric collision sequence $C_{L,j,1}$, $C_{R,j,1}$, $C_{R,j,2}$, $C_{L,j,2}$ with its left ($L$) and right ($R$) bath ancillas; each
ancilla is reset to $\lvert 0\rangle$ before every collision.
(b) Structure of a single collision $C_{\alpha,j,k}$: the ancilla is rotated by $R_y(\eta_{\alpha,j})$, and a projective energy
measurement yields $E^{\mathrm{in}}_{\alpha,j,k}$ while preparing the thermal state. After the partial-swap gate $S_{\alpha,j}$, a second measurement yields $E^{\mathrm{out}}_{\alpha,j,k}$, giving the stochastic heat $Q_{\alpha,j,k}=E^{\mathrm{out}}_{\alpha,j,k}-E^{\mathrm{in}}_{\alpha,j,k}$ delivered to bath $\alpha$. }
    \label{fig:circuit-step}
\end{figure}
We use three qubits to implement this collision model on a digital quantum circuit,
the working qubit and two ancillas. Figure \ref{fig:circuit-step} shows a single cell of the circuit. We represent the working qubit in its instantaneous basis using the rotation $R_j = \exp[-i\vartheta(\overline{t}_j)\hat{\sigma}_q^y/2]$, so that $R^\dagger_j \hat{\tau}^\pm_{q,j}R_j=\hat{\sigma}_q^\pm$. In this encoding, Eq. (\ref{eq:swap}) describes a partial excitation-exchange gate. 
The coherent evolution $U_j$ includes the basis update between successive cells and the precession
generated by the frozen Hamiltonian $B_j\hat\tau_q^z$.

To reduce discretization error and reservoir-order bias, we implement four collisions per cell, which are arranged in the symmetric order
${\rm L-R-R-L}$~\cite{Huyghebaert1990Dec,Poulin2011Apr}.
These are indicated as $C_{\alpha,j,k}$ in Fig.~\ref{fig:circuit-step} (a), with $k=1,2$. 
Before each collision, the ancilla is reset to $\lvert0\rangle$. 
A rotation $R_y(\eta_{\alpha,j})$ with $\eta_{\alpha,j}=2\arccos\sqrt{p^e_{j}}$ followed by a projective energy measurement allows their preparation in the thermal state of Eq. (\ref{eq:thermal}).
A second energy measurement follows the partial swap.
The two outcomes record the ancilla energies before and after the interaction, from which the heat is calculated
below.

{\em Lindbladian dynamics.}
Expanding Eq.~(\ref{eq:swap}) for $\Delta t\to0$ at fixed $\mathcal T$ and tracing the ancillas yields the disipative contribution. Together with the coherent contribution gives the Lindblad equation~\cite{Ciccarello2022Apr,sm}:
\begin{equation}\label{eq:lindblad-0}
\frac{d\rho_q(t)}{dt} = -i \left[  {\cal H}_q(t), \rho_q(t) \right] + \sum_{\alpha={\rm L,R},s=\pm}\Gamma^s_\alpha(t) \mathcal{D}[\hat{\tau}^s_q]\rho_q(t),
\end{equation}
where  $\mathcal{D}[A]\rho = A\rho A^\dagger-\left\{A^\dagger A,\rho \right\}/2$ and the time-dependent excitation/relaxation rates 
\begin{equation}
  \Gamma_\alpha^+(t)= \Gamma_\alpha(t)p^e(t), \qquad \Gamma_\alpha^-(t)= \Gamma_\alpha(t)[1-p^e(t)],
\end{equation}
with  
$p^e(t)=1/\left[1+e^{2 B(t)/T}\right]$,
 satisfy detailed balance.
For frozen parameters, Eq.~(\ref{eq:lindblad-0}) describes precession about $\boldsymbol B$, longitudinal relaxation towards 
$-\tanh[B/T]$ 
at the rate $\Gamma_{\rm tot}=\sum_\alpha\Gamma_\alpha$, and transverse decay at a rate $\Gamma_{\rm tot}/2$. The adiabatic dynamics and geometric heat pumping of this driven model have been studied theoretically in 
Refs.~\cite{Bhandari2020Oct,TerrenAlonso2022Feb,Caselli2026May}.

{\em Heat measurement and energy balance.}
In a single qubit-ancilla collision, the partial swap preserves the energy. As a consequence, the energy gain of the ancilla equals the energy loss of the qubit. This allows us to identify the measured energy difference with the heat delivered to the corresponding reservoir.
The ancilla energy measurements immediately before and after each collision yield
$E_{\alpha,j,k}^{\mathrm{in}}$ and $E_{\alpha,j,k}^{\mathrm{out}}$. 
The stochastic heat exchanged between the ancilla and the qubit is 
$Q_{\alpha,j,k} =E_{\alpha,j,k}^{\mathrm{out}} -E_{\alpha,j,k}^{\mathrm{in}}$.
Hence, the ensemble average of the total heat exchange recorded over a drive cycle is,
\begin{align}\label{calor}
  Q^c_\alpha&\equiv
  \sum_{j=0}^{N-1}\sum_{k=1}^{2}\left\langle
    E_{\alpha,j,k}^{\mathrm{out}}
   -E_{\alpha,j,k}^{\mathrm{in}}\right\rangle.
\end{align}
Here, the brackets denote an average over repeated circuit executions (shots).
Each measured
energy uses the ancilla Hamiltonian frozen in its own time cell. Hence, the
time dependence of the level splitting is included in the sum.

In the continuous time limit, the heat per cycle entering the reservoir $\alpha$ and the net work done by the driving read 
\begin{eqnarray}\label{heat-work}
Q_\alpha&=&-\sum_{s=\pm}\int_0^{\cal T} dt\,
\Gamma_\alpha^s(t)
  \operatorname{Tr}\!\left\{
    {\cal H}_q(t)
    \mathcal D[\hat \tau_q^s(t)]\rho_q(t)
  \right\},\nonumber\\
W_q&=&\int_0^{\cal T} dt\,\operatorname{Tr}\!\left\{
    \dot{\cal H}_q(t)
    \rho_q(t)
  \right\},
\end{eqnarray}

where $Q_\alpha >0$ denotes energy transferred from the working qubit to reservoir $\alpha$. At fixed $\mathcal T$, the average calorimetric heat $Q_\alpha^c$ converges to $Q_\alpha$ as $N\to\infty$.

For a periodic state, $\rho_q({\cal T})=\rho_q(0)$, the internal-energy change vanishes. Energy conservation and the second law at equal reservoir temperatures imply
\begin{equation}\label{1-2-law}
\sum_\alpha Q_{\alpha}=W_q \geq 0.
\end{equation}

{\em Slow-driving limit and geometric response.} This limit corresponds to 
a small ratio between the driven-qubit relaxation time and the period. In  this regime the density matrix of the driven-qubit can be expanded as 
$\rho_q(t)\simeq \rho^{(0)}_q(\boldsymbol{B})+\rho^{(1)}_q(\dot{\boldsymbol{B}})+\rho^{(2)}_q(\dot{\boldsymbol{B}},\ddot{\boldsymbol{B}})$ \cite{Cavina2017Aug}. Substituting in Eq. (\ref{heat-work})
and keeping terms up to the lowest non-vanishing order 
we get \cite{Caselli2026May},
\begin{equation}\label{slow}
\begin{aligned}
   Q_\alpha &=Q^{\rm geo}_\alpha + \frac{{\cal C}_\alpha}{\mathcal{T}}+ {\mathcal{O}}\left(\frac{1}{\mathcal{T}^2}\right),
\end{aligned}  
\end{equation}
where
\begin{equation}
Q^{\rm geo}_{\alpha}=\oint d\boldsymbol{B}\cdot \boldsymbol{\Lambda}_\alpha(\boldsymbol{B}). 
\end{equation}The geometric contributions satisfy $Q_{\rm L}^{\rm geo}=-Q_{\rm R}^{\rm geo}$, while
$\sum_\alpha\mathcal C_\alpha/\mathcal T$ equals the leading dissipated work. Under reversal of the protocol
with the same timing profile, the geometric contribution changes sign, whereas the leading dissipative correction
is unchanged.
We stress that $Q^{\rm geo}_{\alpha}$ is purely geometric  since it depends  on the closed path defined by the driving in the parameter space {\em only}, being independent from the velocity or acceleration at which this path is circulated.
The lowest order finite speed correction reads
\begin{eqnarray}\label{second_order_heat}
\frac{\mathcal{C}_\alpha}{\mathcal{T}}=\int_0^{\cal T} dt \dot{\boldsymbol{B}}\cdot \underline{\boldsymbol{\Omega}}_\alpha(\boldsymbol{B})\cdot\dot{\boldsymbol{B}}.
\end{eqnarray}
The response functions are encoded in the vector field $\boldsymbol{\Lambda}_\alpha(\boldsymbol{B})$
and the tensor $\underline{\boldsymbol{\Omega}}_\alpha(\boldsymbol{B})$. The explicit expressions are presented in the SM \cite{sm}.

For equal couplings $g_L=g_R$, the geometric heat is the flux of the curvature
$\mathcal F_\alpha=\partial_{B_x}\Lambda_{\alpha,z} -\partial_{B_z}\Lambda_{\alpha,x}$ through the oriented area $A$ enclosed by the driving contour:
\begin{equation}\label{geo-pump}
\begin{aligned}
  Q_{\rm L}^{\rm geo}
  &=\iint_A\mathcal F_L(B_x,B_z)\,dB_x\,dB_z, \\
  \mathcal F_L&=-\frac{\sin(2\vartheta)}{ T}
             \operatorname{sech}^{2}(B/T).
\end{aligned}
\end{equation}

We consider the following cyclic protocol,
\begin{align}
  B(t)&=B_0+\delta_B\,f[\cos(2 \pi t/\mathcal{T})], \nonumber \\
  \vartheta(t)&=\vartheta_0+\delta_\vartheta\,f[\sin(2 \pi t/\mathcal{T})],
  \label{eq:driving-loop}
\end{align}
where at $\delta_B=0$ the enclosed area is zero and the geometric pumped heat vanishes. For parameters such that
as $A$ tends to fully cover the first quadrant, it approaches the bound $T \ln 2$ 
\cite{TerrenAlonso2022Feb,Caselli2026May}, which coincides with the celebrated Landauer
limit \cite{Landauer1961Jul,Landauer1988Oct}. 

\begin{figure}[t]
  \centering
  \includegraphics[width=\linewidth]{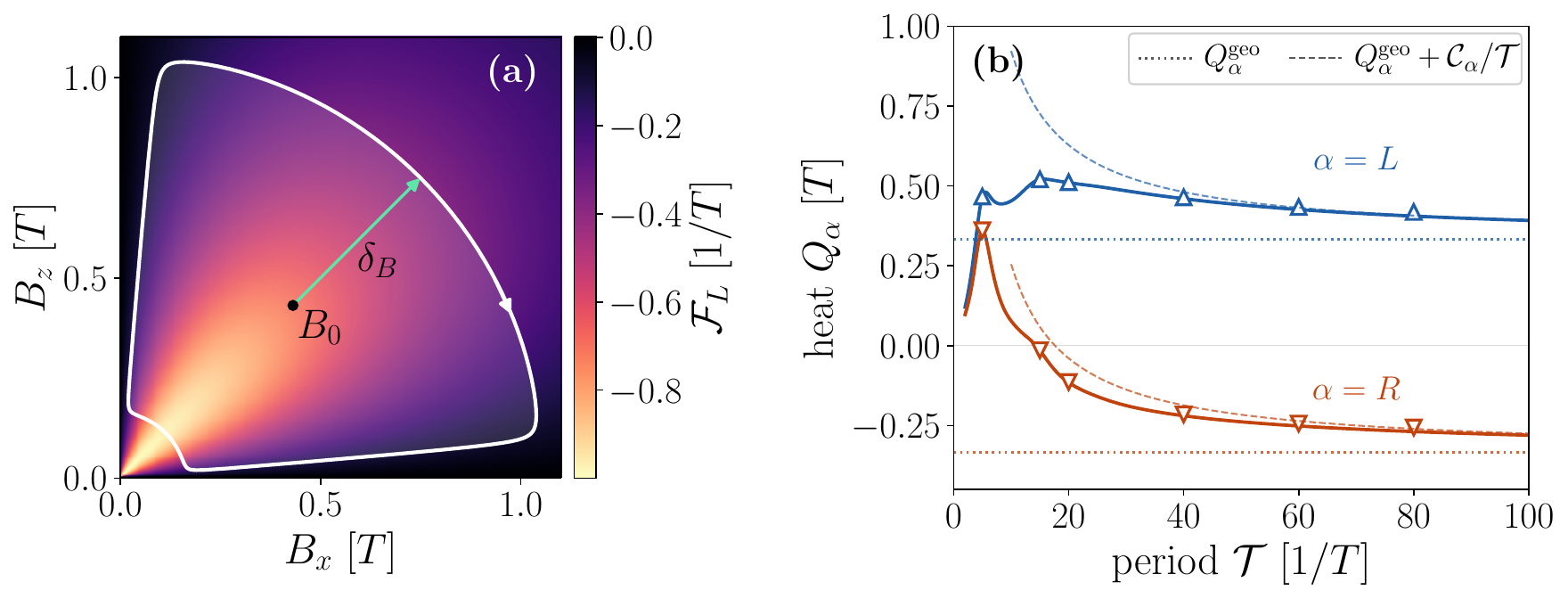}
  \caption{(a) Curvature ${\mathcal{F}_L}$ of the adiabatic pump in the $(B_x,B_z)$ plane and a driving loop of amplitude $\delta_B=0.45$ around $B_0$ ($B_0=0.61$, $\vartheta_0=45^\circ$, $\delta_\vartheta=40^\circ$; 
  $T=g_L=g_R=1$, 
  $\omega_c=120$, $\kappa=2.5$). 
(b)~Heat exchanged with the two reservoirs per driving cycle, for a clockwise sense of traversal, as a function of the period $\mathcal{T}$; blue and orange denote the left ($L$) and right ($R$) reservoirs; positive $Q$ is heat delivered to the reservoir.
Solid lines are the full Lindblad evolution, and open triangles are the results of the collision model with $N=48$ cells per cycle.
Dotted lines are the adiabatic limits $Q^{\rm geo}_{L}=-Q^{\rm geo}_{R}\simeq 0.334$,
and dashed lines are the expansion $Q^{\rm geo}_{\alpha}+\mathcal{C}_\alpha/\mathcal{T}$ with
$\mathcal{C}_L=\mathcal{C}_R\simeq 5.887$.}
  \label{fig:pump0}
\end{figure}

As discussed before, this is not the only component of the heat. Within the slow-driving regime we can solve Lindblad equation by means of 
the adiabatic expansion and calculate explicitly
the  component $\mathcal{C}_\alpha/\mathcal{T}$ 
defined in Eq. (\ref{slow}). We have verified that it satisfies
 ${\cal C}_{\rm L}={\cal C}_{\rm R}={\cal T} W_q/2 >0$ for this configuration. Reversing the traversal with the same timing profile changes the sign of $Q^{\rm geo}_\alpha$ and leaves this quadratic correction unchanged. All these features are consistent with identifying 
 $Q^{\rm geo}_\alpha$ with pumping and ${\cal C}_\alpha$
  with dissipation.

The result of both contributions, $Q_\alpha^{\rm geo}+\mathcal{C}_\alpha/\mathcal{T}$ is shown in dashed lines in Fig. \ref{fig:pump0} (b).
For faster protocols, both, the dissipated and pumped heat are expected to have other components. Therefore, to calculate the heat at each reservoir we must solve the full time-dependent Lindblad equation Eq.~(\ref{eq:lindblad-0}) (see SM \cite{sm} for details). The results are shown in solid lines in Fig. \ref{fig:pump0} (b), along with the solution of the collision model (symbols)~\psc{\cite{qiskit2024}}. 
We find an excellent agreement between both approaches for a number of cells as low as $N=48$ and also  with the slow-driving expansion for long enough periods. Importantly, these results show that for fast cycles (${\cal T} < 14$ for the parameters of the figure), $Q_\alpha>0$ for both reservoirs, which means that the dissipated heat is larger than the pumped component. At short periods, the heat exhibits a nonmonotonic dependence on $\mathcal T$ near the precession timescale
$\pi/B_0$.

Reversing the driving protocol with the same timing profile changes the sign of the geometric heat while leaving the
leading dissipative correction unchanged. We therefore identify
the orientation-odd heat with the pumping component
\begin{equation}\label{eq:oddpumpedheat}
Q_{\alpha,\mathrm{odd}}
=\frac{Q_\alpha^{\cwopencirclearrow}
      -Q_\alpha^{\acwopencirclearrow}}{2},
\end{equation}
where the arrows denote opposite traversal directions of the same contour. In the slow-driving limit of the
continuous dynamics, this quantity approaches $Q_\alpha^{\rm geo}$.

{\em Experimental results.}
We implemented the collision model quantum circuit using qiskit~\cite{qiskit2024} on IBM's Heron r2 processor. 
We ran the circuits on qubits $(q,a_{\rm L},a_{\rm R})=(47,48,46)$ of \texttt{ibm\_kingston}, without error mitigation or readout correction~\cite{sm}. Each circuit contains 384 CZ gates and has an execution time of  $\approx690~\mu$s.
The qubit initial state is prepared in the time-periodic state at time $t=0$, which is obtained with the collision model for each parameter set and traversal direction. 
 Details are shown in the SM~\cite{sm}. 
We execute both traversal directions and obtain the mean calorimetric heat per cycle from Eq.~(\ref{calor}).
  On the processor, noise produces a predominantly orientation-even heat background, substantially larger than the odd pumped heat. Circuit simulations including noise from the processor calibration data, allow to identify contributions from gate errors and relaxation and dephasing during circuit execution. 
The relaxation of the driven qubit during the waiting intervals associated with ancilla measurement and reset has a predominant contribution to the background heat. 
\begin{figure}[t]
  \centering
  \includegraphics[width=\linewidth]{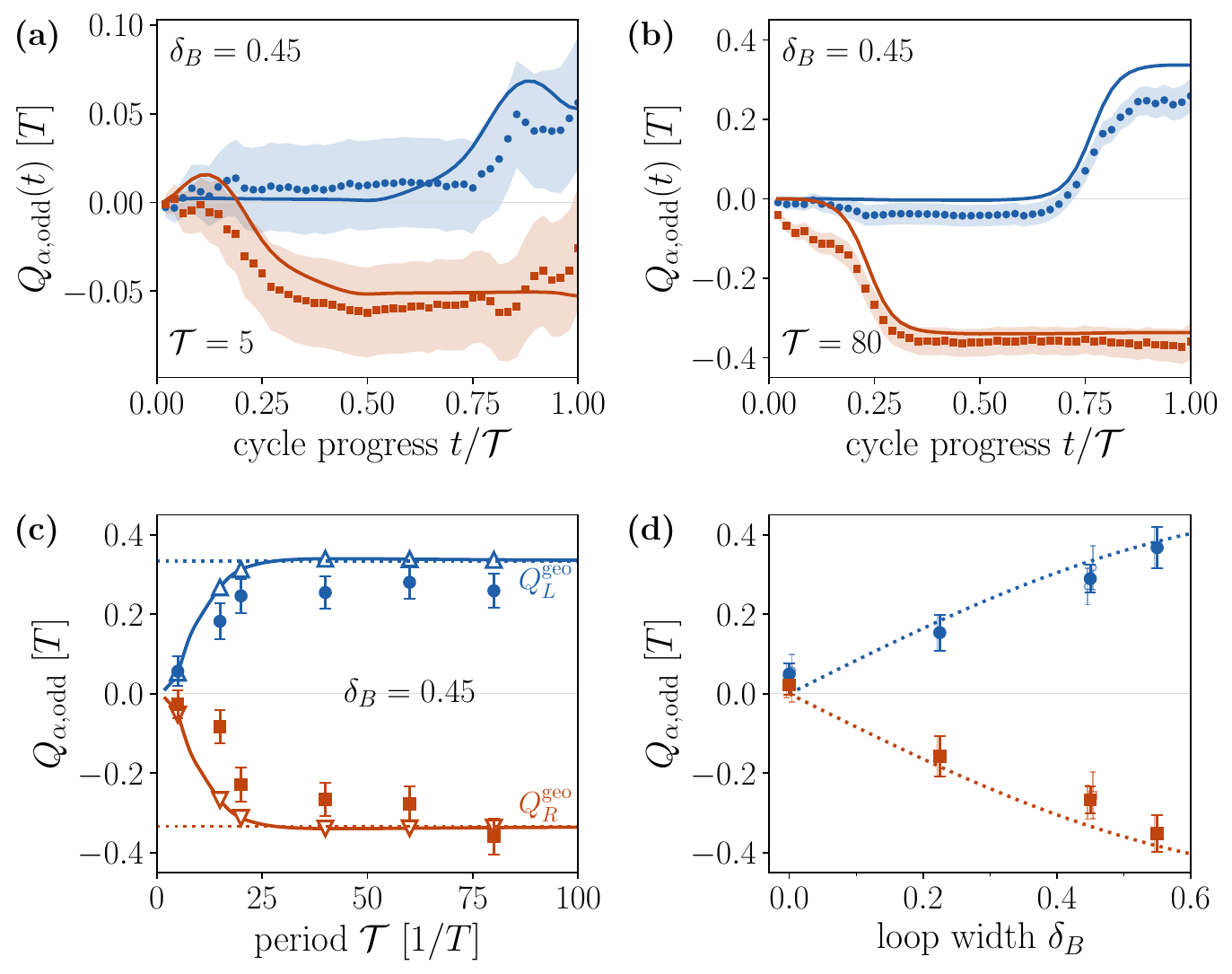}
\caption{Orientation-odd (pumped) heat on \texttt{ibm\_kingston} compared with the exact discrete collision model.
(a)--(b)~Cumulative odd heat $Q_{\alpha,\mathrm{odd}}(t)$ [see Eq. \ref{eq:oddpumpedheat}] exchanged with the left ($L$, blue disks) and right ($R$, orange squares) reservoirs along one cycle, for periods ${\cal T}=5$, and $80$ at 
$\delta_B=0.45$. Lines: ideal collision model with $N=48$ and the same initial state as the measurements; symbols: measurements; shaded bands: $1\sigma$ shot-noise uncertainty.
(c)~Pumped heat $Q_{\alpha,\mathrm{odd}}$ (see Eq.~\ref{eq:oddpumpedheat}). 
Filled symbols with error bars are the quantum processor measurements ($\pm1\sigma$). 
The odd heat components have opposite signs $Q_{L,\mathrm{odd}}\simeq -Q_{R,\mathrm{odd}}$, which is consistent with the predicted pumping response.
(d)~Pumped heat per cycle $Q_{\alpha,\mathrm{odd}}$ versus loop width $\delta_B$ for ${\cal T}=60$; error bars: $1\sigma$ shot noise.
Measurements use $3000$ shots per circuit, $N=48$ collision cells per period, with no error mitigation. Loop parameters as in Fig. \ref{fig:pump0}.}
\label{fig:pump1}
\end{figure}

The quantum processor measurements are presented in Fig.~\ref{fig:pump1} against the theoretical predictions. 
The measured cumulative pumped heat along the cycle shown in panels (a)-(b) for two driving regimes, $\mathcal T=5$ and  $\mathcal T=80$, respectively, displays a remarkable agreement with the theoretical calculations. This is also the case for net pumped component for cycles with a fixed value of $\delta_B$ 
parameter of the protocol Eq. (\ref{eq:driving-loop}) shown in panel (c),  as a function of the period. 
In the adiabatic regime, the heat converges to a value smaller in absolute value than the expected geometric value (shown in dashed lines). The source of this discrepancy can be traced back to the quantum processor gate errors and relaxation times (see SM \cite{sm}). The half-difference in Eq.~(\ref{eq:oddpumpedheat}) cancels the orientation-even component of this background, while a much smaller residual odd contribution attenuates the pumping signal.
Results for a period ${\cal T}=60$ (deep in the adiabatic regime) and different 
values of $\delta_B$  are shown in the panel (d).  
As expected, the odd heat grows with the area of the loop and vanishes within errors at $\delta_B=0$, where the area is zero.

{\em Summary and conclusions.} We have demonstrated geometric heat pumping on a
quantum processor by cyclically driving a working qubit asymmetrically
coupled to two ancillas. Repeated ancilla interactions emulate thermal
reservoirs and  a net energy transfer between them takes place in the cycle. Although we
focused on equal reservoir temperatures, the pumping remains robust against
a small thermal bias in the numerical simulations in the intermediate $\mathcal{T}$ regime (not shown). 

In the present implementation, noise-induced heating exceeds the pumped heat, precluding net cooling.
Although Eq.~(\ref{eq:oddpumpedheat}) cancels the orientation-even background in the measured estimator, it does not remove the physical heating. 
Circuit-level noise simulations identify gate errors and relaxation
and dephasing as important contributions, with the working qubit particularly affected during the waiting
intervals associated with ancilla measurement and reset.
Reducing these waiting times, improving gate fidelities, and increasing coherence times are therefore promising routes towards net heat extraction.
An interesting direction is using more than one working qubit to generate the pumping \cite{Caselli2026May}.

These results establish a proof of concept for geometric control of
 heat flow, and ultimately cooling qubits, through programmable driving protocols.

\textit{Acknowledgments}--- 
We acknowledge the use of IBM Quantum services for this work. The views expressed are those of the authors, and do not reflect the official policy or position of IBM or the IBM Quantum team. We acknowledge support from CONICET and from Simons-Balseiro program (GC). We thank Leandro Tosi for a critical reading of the manuscript.

\textit{Data availability}---The data and code that support the findings of this article are openly available~\cite{zenodo-geopump}.

\bibliography{bibliography.bib,collisions.bib}

\end{document}


\title{Supplemental Material for ``Geometric heat pumping on a quantum processor''}

\author{Ger\'onimo Caselli~\orcidlink{.....}}
\affiliation{Facultad de Ciencias Exactas, Ingenier\'ia y Agrimensura, Universidad Nacional de Rosario and Instituto de F\'isica Rosario (CONICET), Rosario, Argentina}
\author{Luis O. Manuel~\orcidlink{.....}}
\affiliation{Facultad de Ciencias Exactas, Ingenier\'ia y Agrimensura, Universidad Nacional de Rosario and Instituto de F\'isica Rosario (CONICET), Rosario, Argentina}
\author{Liliana Arrachea~\orcidlink{0000-0002-7223-4610}}
\affiliation{Centro At\'omico Bariloche, Instituto de Nanociencia y Nanotecnolog\'ia CONICET-CNEA and Instituto Balseiro (8400), San Carlos de Bariloche, Argentina}
\author{Pablo S. Cornaglia~\orcidlink{0000-0003-4991-6573}}
\affiliation{Centro At\'omico Bariloche, Instituto de Nanociencia y Nanotecnolog\'ia CONICET-CNEA and Instituto Balseiro (8400), San Carlos de Bariloche, Argentina}

\date{\today}
\maketitle

\tableofcontents
\clearpage

\section{Implementation of the single-qubit geometric heat pump on an IBM quantum processor}

This Supplemental Material describes how one cycle of the geometric heat pump is implemented on
IBM superconducting processors of the Heron~r2 generation. The data of the main text were taken on
\code{ibm\_kingston}. We use $\hbar=k_B=1$ and express energies in units of the reservoir
temperature $T$.

\subsection{Model and driving protocol}

The working qubit $q$ has the Hamiltonian
\begin{equation}
  \mathcal H_q(t)= \bm{B}(t) \cdot \hat{\bm{\sigma}}_q=B(t)\,\hat\tau^z_q(t),\qquad
  \bm B(t)=B(t)\,\bm n_B(t),\qquad \bm n_B(t)=\bigl(\sin\vartheta(t),0,\cos\vartheta(t)\bigr),
  \label{eq:H}
\end{equation}
where $\hat{\bm\sigma}_q$ are Pauli matrices, so that the instantaneous energies are $\pm B$. The
operators $\hat\tau^{z,\pm}_q(t)=U(t)\,\hat\sigma^{z,\pm}_q\,U^\dagger(t)$, with
$U(t)=e^{-i\vartheta(t)\hat\sigma^y_q/2}$, are the Pauli and ladder operators along $\bm n_B(t)$.
The field follows the closed loop
\begin{equation}
  B(t)=B_0+\delta_B\,f\!\left[\cos\phi(t)\right],\qquad
  \vartheta(t)=\vartheta_0+\delta_\vartheta\,f\!\left[\sin\phi(t)\right],\qquad
  f(x)=\frac{\tanh(\kappa x)}{\tanh \kappa}.
  \label{eq:loop}
\end{equation}
For $\kappa\to0$ the loop is an ellipse in the $(B,\vartheta)$ plane; for large $\kappa$ it approaches the
rectangle $[B_0-\delta_B,B_0+\delta_B]\times[\vartheta_0-\delta_\vartheta,\vartheta_0+\delta_\vartheta]$, so that for
$\kappa=2.5$ it is a rectangle with rounded corners (``squircle''). In the $(B_x,B_z)$ plane it is a
sector of an annulus with rounded corners. The loop is traversed in the two directions
\begin{equation}
  \phi^{\fwd}(t)=2\pi\,\frac{t}{\T},\qquad \phi^{\bwd}(t)=2\pi\left(1-\frac{t}{\T}\right),
  \label{eq:orientation}
\end{equation}
both starting from the same point. For $\delta_B>0$ the forward direction $\fwd$ is clockwise in the
$(B_x,B_z)$ plane (with $B_x$ on the horizontal axis). Changing the sign of $\delta_B$ reverses the
direction. For $\delta_B=0$ the loop encloses no area.

The qubit is coupled to two Caldeira--Leggett Ohmic reservoirs, left ($L$) and right ($R$), both at
temperature $T$ and with cutoff frequency $\omega_c$, through the operators
$g_\alpha\,\bm {n}_\alpha\cdot\hat{\bm{\sigma}}_q$, with $\bm{n}_L=\hat{\bm x}$ and $\bm{n}_R=\hat{\bm z}$, and
with the spectral densities normalized such that the rates are those of Eq.~\eqref{eq:rates_pm}. Only
the component of $\bm{n}_\alpha$ transverse to $\bm{n}_B$ induces transitions, which gives the factor
$|\bm{n}_\alpha\times\bm{n}_B|^2$, i.e.\ $\cos^2\vartheta$ for $L$ and $\sin^2\vartheta$ for $R$. The
excitation ($+$) and relaxation ($-$) rates are
\begin{equation}
  \Gamma^\pm_\alpha(t)=\frac{\gamma[B(t)]}{2\left[e^{\pm2B(t)/T}+1\right]}\,g_\alpha^2\,
  |\bm n_\alpha\times\bm n_B(t)|^2,
  \qquad
  \gamma[B]=\frac{B}{\tanh(B/T)}\,e^{-2B/\omega_c},\qquad \alpha=L,R,
  \label{eq:rates_pm}
\end{equation}
and the population relaxation rate of each reservoir is
\begin{equation}
  \Gamma_\alpha=\Gamma^+_\alpha+\Gamma^-_\alpha=\tfrac12\,\gamma[B]\,g_\alpha^2\,|\bm n_\alpha\times\bm n_B|^2,
  \qquad
  \Gamma_L=\tfrac12 g_L^2\cos^2\vartheta\,\gamma[B],\quad
  \Gamma_R=\tfrac12 g_R^2\sin^2\vartheta\,\gamma[B] .
  \label{eq:rates}
\end{equation}
The equilibrium polarization along the field is $z_{\rm eq}=-\tanh(B/T)$. 


\subsection{Collision-model discretization}
\label{sec:discrete}

The period is divided into $N$ cells of equal duration,
\begin{equation}
  \Delta t=\frac{\T}{N},\qquad t_j=j\,\Delta t,\qquad \bar t_j=t_j+\frac{\Delta t}{2},\qquad j=0,\dots,N-1.
\end{equation}
All parameters of cell $j$ are evaluated at its midpoint along the respective direction of the loop:
$B_j=B(\bar t_j)$, $\vartheta_j=\vartheta(\bar t_j)$ and $\Gamma_{\alpha,j}=\Gamma_\alpha(\bar t_j)$.
(The loop centre $B_0$, $\vartheta_0$ of Eq.~\eqref{eq:loop} is not the value of cell $0$, which we write
$B(\bar t_0)$, $\vartheta(\bar t_0)$.) The sequence of cells of direction $\bwd$ is the sequence of
direction $\fwd$ in reverse order.

Within a cell the qubit evolves under the frozen Hamiltonian $B_j\hat\tau^z_q$ and interacts with each
reservoir through two collisions, in the symmetric order $L,R,R,L$. Each collision represents half a
cell of the corresponding reservoir, with exchange probability
\begin{equation}
  \lambda_{\alpha,j}=1-e^{-\Gamma_{\alpha,j}\Delta t/2}.
  \label{eq:lambda}
\end{equation}
The symmetric order is a second-order (Strang) splitting of the simultaneous action of both
reservoirs: its deviation from the continuous evolution decreases as $\Delta t^2$, and it treats $L$
and $R$ equivalently. A sequential order $L,R$ would converge only as $\Delta t$ and would assign a
larger share of the dissipated heat to the reservoir acting first in each cell.

A collision with a fresh ancilla in its thermal state acts on the Bloch vector $(x,y,z)$ of $q$ in the
energy frame (Sec.~\ref{sec:circuit}) as a generalized amplitude-damping channel,
\begin{equation}
  (x,y,z)\longmapsto\left(\sqrt{1-\lambda_{\alpha,j}}\,x,\;\sqrt{1-\lambda_{\alpha,j}}\,y,\;
  (1-\lambda_{\alpha,j})\,z+\lambda_{\alpha,j}\,z_{{\rm eq},j}\right),
  \label{eq:gad}
\end{equation}
and deposits into the reservoir the mean heat
\begin{equation}
  \langle Q_{\alpha,j,k}\rangle=B_j\,\lambda_{\alpha,j}\left(z-z_{{\rm eq},j}\right),
  \qquad k=1,2,
  \label{eq:qmean}
\end{equation}
where $z$ is the polarization immediately before the collision and $k$ labels the two collisions
of reservoir $\alpha$ in cell $j$. Heat is counted positive when it enters the reservoir.
For $N\to\infty$ at fixed $\T$ the cycle map converges to the continuous Lindblad evolution of
Sec.~\ref{full_lindblad}.

\subsection{Quantum circuit}
\label{sec:circuit}

\subsubsection{Qubits and energy frame}

The circuit uses three physical qubits: the working qubit $q$ and two ancillas $a_L$, $a_R$, one per
reservoir, each coupled directly to $q$. The state of $q$ is stored in the instantaneous energy
frame, defined by $\mathcal H_q=U\,B\hat\sigma^z_q\,U^\dagger$, with $U=R_y(\vartheta)$ and
$R_y(\beta)=e^{-i\beta\hat\sigma^y/2}$. In this frame $\hat\tau^{z,\pm}_q$ act as
$\hat\sigma^{z,\pm}_q$, the computational state $|0\rangle$ is the excited level ($+B$) and
$|1\rangle$ the ground level ($-B$). The same convention holds for each ancilla, whose Hamiltonian
during cell $j$ is $B_j\hat\tau^z_\alpha$: a measured bit $b\in\{0,1\}$ corresponds to the energy
$E(b)=B_j(1-2b)$.

\subsubsection{Gates of one cell}

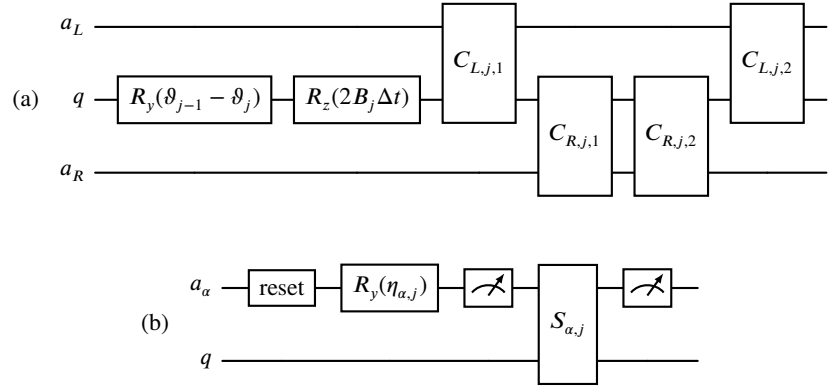
\begin{figure}[t]
\centering
(a)\;
\begin{quantikz}[row sep=0.35cm, column sep=0.3cm]
\lstick{$a_L$} & & & \gate[2]{C_{L,j,1}} & & & \gate[2]{C_{L,j,2}} & \\
\lstick{$q$} & \gate{R_y(\vartheta_{j-1}-\vartheta_j)} & \gate{R_z(2B_j\Delta t)} & &
  \gate[2]{C_{R,j,1}} & \gate[2]{C_{R,j,2}} & & \\
\lstick{$a_R$} & & & & & & &
\end{quantikz}
\\[0.6cm]
(b)\;
\begin{quantikz}[row sep=0.35cm, column sep=0.35cm]
\lstick{$a_\alpha$} & \gate{\text{reset}} & \gate{R_y(\eta_{\alpha,j})} & \meter{} &
  \gate[2]{S_{\alpha,j}} & \meter{} & \\
\lstick{$q$} & & & & & &
\end{quantikz}
\caption{(a) Gates of cell $j$. The first two gates form the coherent step $U_j$; the frame update
is omitted for $j=0$. (b) One collision $C_{\alpha,j,k}$. The first measurement records $b^{\rm in}$
and completes the thermal preparation; the second records $b^{\rm out}$. $S_{\alpha,j}$ is the partial
swap of Eq.~\eqref{eq:swap}, compiled into two CZ gates and single-qubit rotations.}
\label{fig:circuit}
\end{figure}

Cell $j$ consists of the following operations (Fig.~\ref{fig:circuit}):
\begin{enumerate}
\item Coherent step $U_j=R_z(2B_j\Delta t)\,R_y(\vartheta_{j-1}-\vartheta_j)$ on $q$, with
  $R_z(\chi)=e^{-i\chi\hat\sigma^z/2}$. The rotation $R_y(\vartheta_{j-1}-\vartheta_j)$ (omitted for $j=0$)
  re-expresses the state in the energy frame of the new cell, and $R_z(2B_j\Delta t)=e^{-iB_j\hat\sigma^z_q\Delta t}$
  is the precession under the frozen Hamiltonian.
\item Four collisions $C_{L,j,1}$, $C_{R,j,1}$, $C_{R,j,2}$, $C_{L,j,2}$.
\end{enumerate}
Each collision $C_{\alpha,j,k}$ with ancilla $a_\alpha$ is
\begin{enumerate}
\item reset of $a_\alpha$ to $|0\rangle$;
\item rotation $R_y(\eta_{\alpha,j})$ of $a_\alpha$, with
  \begin{equation}
    \eta_{\alpha,j}=2\arccos\sqrt{p^e_{\alpha,j}},\qquad p^e_{\alpha,j}=\frac{1}{1+e^{2B_j/T}},
  \end{equation}
  so that $a_\alpha$ is found in the excited level $|0\rangle$ with probability $p^e_{\alpha,j}$;
\item energy measurement of $a_\alpha$, giving the incoming bit $b^{\rm in}$;
\item partial swap $S_{\alpha,j}$ on $(q,a_\alpha)$;
\item energy measurement of $a_\alpha$, giving the outgoing bit $b^{\rm out}$.
\end{enumerate}
Averaged over its outcome, the first measurement turns the coherent superposition into the thermal
state $\rho_\alpha(t^0_j)=p^e_{\alpha,j}|0\rangle\langle0|+(1-p^e_{\alpha,j})|1\rangle\langle1|$, and at
the same time records the incoming energy. No reset is applied between this measurement and the
interaction. The partial swap is
\begin{equation}
  S_{\alpha,j}=\exp\!\left[-i\varphi_{\alpha,j}\left(\hat\tau^+_q\hat\tau^-_\alpha
  +\hat\tau^-_q\hat\tau^+_\alpha\right)\right],\qquad
  \cos^2\varphi_{\alpha,j}=e^{-\Gamma_{\alpha,j}\Delta t/2}=1-\lambda_{\alpha,j},
  \label{eq:swap}
\end{equation}
implemented by the Qiskit gate \code{XXPlusYYGate}$(2\varphi_{\alpha,j},\,\beta{=}0)$. It exchanges
one excitation between $q$ and $a_\alpha$ with probability $\lambda_{\alpha,j}$. Tracing out the
ancilla gives exactly the channel of Eq.~\eqref{eq:gad}. The working qubit is not measured during
the cycle.

After the last cell, $R_y\bigl(\vartheta_{N-1}-\vartheta(\bar t_0)\bigr)$ returns $q$ to the frame of
the first cell, where the initial state was defined, and $q$ is measured once. One circuit therefore
runs exactly one cycle.

\subsubsection{Classical registers and resources}

Each circuit writes to three classical registers. The register $e_L$ ($e_R$) has $4N$ bits: the
$\ell$-th collision with $a_L$ ($a_R$), $\ell=0,\dots,2N-1$ in time order, writes $b^{\rm in}$ to
bit $2\ell$ and $b^{\rm out}$ to bit $2\ell+1$. A one-bit register holds the final bit of $q$. For
$N=48$ a circuit contains $4N=192$ ancilla resets, $8N+1=385$ measurements and $2N=96$ partial swaps
per ancilla, i.e.\ 192 partial swaps in total.

\subsection{Initial state: periodic state of the discrete map}
\label{sec:initial}

The heat is measured in the periodic regime, without running transient cycles on hardware. For a
given direction, the full cycle of the discrete model (Sec.~\ref{sec:discrete}, including the closing
frame rotation) maps the energy-frame Bloch vector affinely,
\begin{equation}
  \bm v_{\rm end}=A\,\bm v_0+\bm b .
\end{equation}
The vector $\bm b$ is obtained by propagating $\bm v_0=\bm 0$ and column $i$ of $A$ by
propagating the unit vector $\bm e_i$ and subtracting $\bm b$. The periodic state solves
$(I-A)\,\bm v_*=\bm b$. It is computed separately for each direction, $\delta_B$, $\T$ and $N$.

A mixed qubit state is prepared as an ensemble of its two eigenstates. With $r=|\bm v_*|$ and
$\bm n=\bm v_*/r$,
\begin{equation}
  \rho_q^*=\sum_{s=\pm}p_s\,|s\bm n\rangle\langle s\bm n|,\qquad
  p_\pm=\frac{1\pm r}{2}.
\end{equation}
Although $\rho_q^*$ has the form of a Gibbs state of $-\bm n\cdot\hat{\bm\sigma}_q$ at an effective
temperature, it is not the thermal state of $\mathcal H_q(t_0)$. First, $\bm n$ is tilted away from the
field axis (the $z$ axis of the energy frame) and has a $y$ component, i.e.\ $\rho_q^*$ has coherences
in the energy basis. Second, the weights $p_\pm$ differ from the Boltzmann weights at temperature $T$.
Both differences vanish only for
$\T\to\infty$ and $N\to\infty$. At finite $N$ a tilt $\vartheta_{N-1}-\vartheta(\bar t_0)$ remains even for
$\T\to\infty$, because the state defined in the frame of the first cell has relaxed during the last
cell of the previous cycle.

Each pure component $s\bm n=(n_x,n_y,n_z)$ is prepared from $|0\rangle$ by $R_y(\arccos n_z)$ followed
by $R_z(\operatorname{atan2}(n_y,n_x))$ and is run as a separate circuit. A parameter point therefore
consists of four circuits,
\begin{center}
\begin{tabular}{lll}
\toprule
Direction & Initial component & Weight\\\midrule
$\fwd$ & $+\bm n^{\fwd}$, $-\bm n^{\fwd}$ & $p^{\fwd}_+$, $p^{\fwd}_-$\\
$\bwd$ & $+\bm n^{\bwd}$, $-\bm n^{\bwd}$ & $p^{\bwd}_+$, $p^{\bwd}_-$\\
\bottomrule
\end{tabular}
\end{center}
all run with the same number of shots. The weights enter only in the analysis
(Sec.~\ref{sec:estimators}). By linearity of the quantum evolution and of the measurement
statistics, the weighted combination of the two components reproduces the statistics of the mixed
initial state $\rho_q^*$.

\subsection{Heat reconstruction}
\label{sec:estimators}

\subsubsection{Heat per shot}

The heat delivered to reservoir $\alpha$ in one collision is the change of the ancilla energy,
\begin{equation}
  Q_{\alpha,j,k}=E^{\rm out}_{\alpha,j,k}-E^{\rm in}_{\alpha,j,k}
  =2B_j\left(b^{\rm in}_{\alpha,j,k}-b^{\rm out}_{\alpha,j,k}\right)\in\{0,\pm2B_j\}.
\end{equation}
Each energy uses the ancilla splitting of its own cell. For every shot we form the heat per cell,
$Q_{\alpha,j}=\sum_{k=1}^{2}Q_{\alpha,j,k}$, and the heat per cycle,
$Q_\alpha=\sum_{j=0}^{N-1}Q_{\alpha,j}$. Its average over shots equals the sum of
Eq.~\eqref{eq:qmean} over all collisions. The records of one shot are kept together: successive
collisions of the same shot are not treated as independent samples.

\subsubsection{Estimators and uncertainties}

For direction $d\in\{\fwd,\bwd\}$ and initial component $s=\pm$, let $\bar Q^{d}_{\alpha,s}$ and
$v^{d}_{\alpha,s}$ be the sample mean and sample variance of the per-shot heat over the $n^d_s$
shots, and $p^d_s$ the weight of the component. The heat of direction $d$ and its variance are
\begin{equation}
  Q^{d}_\alpha=\sum_{s=\pm} p^d_s\,\bar Q^{d}_{\alpha,s},\qquad
  V^{d}_\alpha=\sum_{s=\pm}\frac{(p^d_s)^2\,v^{d}_{\alpha,s}}{n^d_s}.
\end{equation}
The orientation-odd (pumped) and orientation-even parts are
\begin{equation}
  Q_{\alpha,\odd}=\frac{Q^{\fwd}_\alpha-Q^{\bwd}_\alpha}{2},\qquad
  Q_{\alpha,\even}=\frac{Q^{\fwd}_\alpha+Q^{\bwd}_\alpha}{2},
\end{equation}
both with variance $\bigl(V^{\fwd}_\alpha+V^{\bwd}_\alpha\bigr)/4$, since the two directions are
run in independent circuits. The $L$ and $R$ records of the same shot are correlated. Their covariance
$C^{d}=\sum_s (p^d_s)^2\operatorname{Cov}^d_s(Q_L,Q_R)/n^d_s$ enters any combination of both
reservoirs, for instance
$\operatorname{Var}(Q_{L,\odd}+Q_{R,\odd})=\operatorname{Var}Q_{L,\odd}+\operatorname{Var}Q_{R,\odd}
+2C_\odd$, with $C_\odd=\bigl(C^{\fwd}+C^{\bwd}\bigr)/4$. The error bars are one standard error of
these estimators, i.e.\ shot noise only.

\subsubsection{Cumulative heat along the cycle}

To follow the heat along the cycle, the estimators above are applied to the per-shot partial sums
over cells. The two directions visit the loop in opposite order. To compare them on the same
portion of the loop, the cells of direction $\bwd$ are relabelled by their position on the loop
(cell $j$ of direction $\bwd$ sits at the midpoint of cell $N-1-j$ of direction $\fwd$). The
cumulative odd heat after the first $m$ cells of the loop is
\begin{equation}
  Q_{\alpha,\odd}(t_m)=\frac12\left[\sum_{j=0}^{m-1}Q^{\fwd}_{\alpha,j}
  -\sum_{j=N-m}^{N-1}Q^{\bwd}_{\alpha,j}\right],\qquad m=1,\dots,N,
\end{equation}
where the second sum runs over the last $m$ cells executed in direction $\bwd$. For $m=N$ it
reduces to $Q_{\alpha,\odd}$.

\subsection{Reference calculations and circuit validation}
\label{sec:references}

\subsubsection{Validation of the circuits}

Before execution, the circuits of every parameter point are checked against the discrete model in
two independent ways.
\emph{Exact gate-level propagation:} the density matrix of $q$ is propagated through the gates of
each circuit, using the matrix of the Qiskit \code{XXPlusYYGate} acting on $q$ and a fresh
thermal ancilla, followed by the partial trace. The ancilla populations before and after each
collision give the expected heat of every collision. The weighted heats per cell, the cumulative
odd heats and the final population of $q$ agree with the discrete model to within $10^{-13}$.
\emph{Noiseless sampling:} the circuits are sampled on the Qiskit Aer simulator and processed with
the same analysis as the hardware data. The estimated heats agree with the discrete model within
shot noise.

\subsection{Execution on IBM Heron r2 processors}

The circuit uses only the native operations of IBM Heron~r2 processors: CZ, $R_z$, $\sqrt X$,
mid-circuit measurement and reset. It needs three physical qubits, with $q$ coupled directly to both
ancillas.
\begin{itemize}
\item \emph{Physical qubits.} The data of the main text were taken on \code{ibm\_kingston} with
  $(q,a_L,a_R)=(47,48,46)$.
\item \emph{Compilation.} Qiskit preset pass manager with \code{optimization\_level=3},
  \code{seed\_transpiler=0} and the chosen triple as fixed initial layout. After compilation it is
  checked that every measurement of $q$, $a_L$ and $a_R$ writes into its own register. Each partial
  swap compiles into two CZ gates, giving 384 CZ gates per circuit for $N=48$. On
  \code{ibm\_kingston} the compiled depth is about 2590 and the scheduled duration of one circuit
  about $0.69$~ms.
\item \emph{Execution.} The four compiled circuits of a parameter point are submitted together as
  one \code{SamplerV2} job with 3000 shots per circuit and the default runtime options. No error
  mitigation, dynamical decoupling or twirling is applied, and no readout correction is made.
\end{itemize}

\subsection{Effect of hardware noise} \label{sec:noise}

Hardware noise adds to the measured heats a contribution $\Delta Q^{c}_\alpha=Q^{c}_\alpha-Q^{c}_{\alpha,\rm ideal}$, where $Q^{c}_{\alpha,\rm ideal}$ is the prediction of the noiseless collision model. 
Table~\ref{tab:noise} shows its even and odd parts, $\Delta Q_{\alpha,\even}=(\Delta Q^{\fwd}_\alpha+\Delta Q^{\bwd}_\alpha)/2$ and $\Delta Q_{\alpha,\odd}=(\Delta Q^{\fwd}_\alpha-\Delta Q^{\bwd}_\alpha)/2$, for the data of the main
text. The even part is 1--5 $ T$ per cycle, several times larger than $|Q^{(\rm geo)}_\alpha|\simeq0.34\,  T$.
The odd part is below 3\% of the even part for $\T\ge40$ and about 10\% for $\T\le20$. The
noise contribution is therefore predominantly direction-symmetric, and it cancels in
$Q_{\alpha,\rm{odd}}$.

To identify its origin, the submitted circuits are simulated with Qiskit Aer and a noise model with typical parameters of IBM Heron~r2 processors. 
The model includes readout assignment errors $P(0|1)=1.2\times10^{-2}$ and $P(1|0)=4\times10^{-3}$, a reset error of $10^{-3}$, CZ errors of $3.5\times10^{-3}$ (two-qubit depolarizing channel) and $\sqrt X$ errors of $2.3\times10^{-4}$ (single-qubit depolarizing channel). It also includes thermal relaxation with $T_1=160~\mu$s and $T_2=100~\mu$s during every gate, measurement, reset and idle interval, with as-soon-as-possible timing and durations of 30~ns ($\sqrt X$), 80~ns (CZ), 2.1~$\mu$s (measurement) and 2.2~$\mu$s (reset). 
Without noise, this calculation reproduces the discrete model to within $10^{-12}$.

\begin{table}[h]
\caption{Noise contribution to the heat per cycle, $\delta_B=0.45$, $N=48$: hardware (hw, shot
standard error 0.03--0.05) and simulation with typical Heron~r2 noise parameters (sim). Heat is in units of $ T$.}
\label{tab:noise}
\begin{ruledtabular}
\begin{tabular}{ccccccccc}
 & \multicolumn{2}{c}{$\Delta Q_{L,\even}$} & \multicolumn{2}{c}{$\Delta Q_{L,\odd}$}
 & \multicolumn{2}{c}{$\Delta Q_{R,\even}$} & \multicolumn{2}{c}{$\Delta Q_{R,\odd}$}\\
$\T$ & hw & sim & hw & sim & hw & sim & hw & sim\\
\hline
5 & 1.56 & 0.51 & $+$0.00 & $-$0.04 & 1.12 & 0.51 & $+$0.03 & $+$0.05 \\
15 & 2.00 & 1.07 & $-$0.09 & $-$0.13 & 1.66 & 1.08 & $+$0.18 & $+$0.16 \\
20 & 2.08 & 1.25 & $-$0.07 & $-$0.15 & 1.73 & 1.25 & $+$0.08 & $+$0.16 \\
40 & 3.53 & 3.04 & $-$0.08 & $-$0.11 & 4.42 & 3.05 & $+$0.07 & $+$0.13 \\
60 & 3.57 & 3.21 & $-$0.06 & $-$0.09 & 4.56 & 3.21 & $+$0.06 & $+$0.10 \\
80 & 3.67 & 3.30 & $-$0.08 & $-$0.09 & 4.70 & 3.30 & $-$0.02 & $+$0.09 \\
\end{tabular}
\end{ruledtabular}
\end{table}

The simulation reproduces 70--90\% of the measured even part for $\T\ge40$, and 33--72\% for $\T\le 20$. 
Its odd part is below 5\% of the even part for $\T\ge 40$ and below 15\% for $\T\le 20$. 
Adding noise sources cumulatively shows that readout assignment and reset errors alone produce changes below
0.01 $ T$. Including gate errors increases the even excess to approximately 0.5--0.9 $ T$, while subsequently including
relaxation and dephasing yields the full values in Table~\ref{tab:noise}.

One circuit lasts about 0.7~ms, and $q$ is idle during $\approx$ 90\% of it, waiting while the ancillas are measured and reset.
Its relaxation and dephasing during these intervals account for about 25\% of the even part at $\T=40$. 
Varying the measurement and reset durations between 1.6 and 2.7~$\mu$s, or $T_1$ between
120 and 200~$\mu$s and $T_2$ between 80 and 120~$\mu$s, changes the even part by up to 20\%, while
the ratio of odd to even part stays between 2\% and 5\% for $\T = 40$.

\section{Slow-driving limit and geometric response}
In this Supplement we summarize the derivation of the slow-driving expansion of Refs.~\cite{Cavina2017Aug,Bhandari2020Oct,TerrenAlonso2022Feb} up to second order, following the notation of Ref.~\cite{Caselli2026May}. We recall that the reduced density matrix of the qubit evolves according to the Lindblad equation (we set $\hbar = 1$)
\begin{equation}\label{eq:lindblad-0}
\frac{d\rho_q(t)}{dt} = -i \left[ \mathcal{H}_q(t), \rho_q(t) \right] + \sum_{\alpha={\rm L,R}}\,\sum_{s=\pm}\Gamma^s_\alpha(t)\, \mathcal{D}[\hat{\tau}^s_q]\rho_q(t),
\end{equation}
where $\mathcal{H}_q(t) = \boldsymbol{B}(t)\cdot\hat{\boldsymbol{\sigma}}_q$ is the qubit Hamiltonian, $\hat{\tau}^{+}_q$ ($\hat{\tau}^{-}_q$) is the raising (lowering) operator in the instantaneous eigenbasis of $\mathcal{H}_q(t)$, and $\mathcal{D}[A]\rho = A\rho A^\dagger - \{A^\dagger A,\rho\}/2$. The time-dependent excitation ($s=+$) and relaxation ($s=-$) rates are
\begin{equation}
  \Gamma_\alpha^+(t)= \Gamma_\alpha(t)\,p^e(t), \qquad \Gamma_\alpha^-(t)= \Gamma_\alpha(t)\,[1-p^e(t)],
\end{equation}
with $p^e(t)=1/\left[1+e^{2 B(t)/( T)}\right]$ and $B(t) = |\boldsymbol{B}(t)|$.

In what follows, we write Eq.~\eqref{eq:lindblad-0} in the general form
\begin{equation}\label{lindblad}
    \frac{d\rho}{dt} = \mathcal{L}_t[\rho]
    = -i\big[\mathcal{H}_{\mathrm{S}}(\boldsymbol{X}), \rho\big] + \sum_{\alpha} \mathcal{D}_\alpha[\rho],
\end{equation}
with $\rho \equiv \rho_q$, $\mathcal{H}_{\mathrm{S}} \equiv \mathcal{H}_q$, and dissipators
\begin{equation}\label{dissipator}
    \mathcal{D}_\alpha[\rho] = \sum_{\omega>0} \sum_{s=\pm} \Big( L_{\alpha\omega s}\, \rho\, L^{\dagger}_{\alpha\omega s}
    - \frac{1}{2} \big\{ L^{\dagger}_{\alpha\omega s}L_{\alpha\omega s},\, \rho \big\} \Big).
\end{equation}
The jump operators $L_{\alpha\omega s}$ are defined in the instantaneous eigenbasis of the system Hamiltonian, $\mathcal{H}_{\mathrm{S}}|m\rangle = \epsilon_m|m\rangle$, and describe transitions at the Bohr frequency $\omega$ in which the system absorbs ($s=+$) or releases ($s=-$) an energy $\omega$ from or to reservoir $\alpha$. For the qubit, the only Bohr frequency is $\omega = 2B$, and comparison with Eq.~\eqref{eq:lindblad-0} gives $L_{\alpha,2B,\pm} = \sqrt{\Gamma^\pm_\alpha}\,\hat{\tau}^\pm_q$. Both $\mathcal{H}_{\mathrm{S}}$ and $\mathcal{D}_\alpha$, and hence $\mathcal{L}_t$, depend on time only through the control parameters $\boldsymbol{X}(t) = (B(t), \vartheta(t))$, defined by $\boldsymbol{B}(t) = B(t)\,(\sin\vartheta(t), 0, \cos\vartheta(t))$, which are varied periodically in time with period $\mathcal{T}$.

For each time $t$, we define the \emph{frozen} state $\rho^{(f)}(\boldsymbol{X})$, with $\boldsymbol{X} = \boldsymbol{X}(t)$, as the instantaneous stationary solution of Eq.~\eqref{lindblad}, $\mathcal{L}_t\big[\rho^{(f)}(\boldsymbol{X})\big] = 0$, i.e., the state towards which the system would relax if the parameters were held fixed at $\boldsymbol{X}$.
When the parameters change slowly compared with the relaxation time $t_{\mathrm{rel}}$ of the system, the state closely follows the frozen solution, and the deviations from it can be computed systematically. Inserting the ansatz $\rho(t) = \sum_{n\geq 0} \rho^{(n)}(t)$ into Eq.~\eqref{lindblad} and collecting terms with the same number of time derivatives of the driving protocol, we obtain $\rho^{(0)} \equiv \rho^{(f)}$ at zeroth order and the recursion
\begin{equation}\label{recursion}
    \frac{d \rho^{(n-1)}}{dt} = \mathcal{L}_t\big[\rho^{(n)}\big],
    \qquad n \geq 1.
\end{equation}
Since the dynamics is trace preserving and $\mathrm{Tr}\,\rho^{(f)} = 1$, all corrections $\rho^{(n)}$ with $n \geq 1$ are traceless. The recursion is therefore well defined provided $\mathcal{L}_t$ is invertible on the subspace of traceless operators, which is guaranteed when the stationary state is unique. In what follows, $\mathcal{L}_t^{-1}$ denotes this inverse.

Each iteration introduces one additional time derivative of the driving protocol and, therefore, one additional factor of $t_{\mathrm{rel}}/\mathcal{T}$. The procedure thus generates an expansion in powers of the adiabaticity parameter,
\begin{equation}
    \rho(t) = \rho^{(f)}(\boldsymbol{X}) + \rho^{(1)}(\boldsymbol{X}, \dot{\boldsymbol{X}}) + \rho^{(2)}(\boldsymbol{X}, \dot{\boldsymbol{X}}, \ddot{\boldsymbol{X}}) + \ldots,
    \qquad
    \rho^{(n)} = \mathcal{O}\!\left[\left(\frac{t_{\mathrm{rel}}}{\mathcal{T}}\right)^{n}\right].
\end{equation}
Solving the recursion, the first two corrections read
\begin{align}\label{slow-exp}
    \rho^{(1)}(\boldsymbol{X}, \dot{\boldsymbol{X}}) &= \mathcal{L}_t^{-1}\,\frac{d\rho^{(f)}(\boldsymbol{X})}{dt}
    = \left[\mathcal{L}_t^{-1}\,\partial_{\boldsymbol{X}}\right]\rho^{(f)}(\boldsymbol{X})\cdot\dot{\boldsymbol{X}}, \nonumber \\
    \rho^{(2)}(\boldsymbol{X}, \dot{\boldsymbol{X}}, \ddot{\boldsymbol{X}}) &= \mathcal{L}_t^{-1}\,\frac{d\rho^{(1)}}{dt} \nonumber \\
    &= \dot{\boldsymbol{X}}\cdot\left[\mathcal{L}_t^{-1}\,\partial_{\boldsymbol{X}}\right]^2\rho^{(f)}(\boldsymbol{X})\cdot\dot{\boldsymbol{X}}
    + \left[\mathcal{L}_t^{-1}\right]^2\partial_{\boldsymbol{X}}\rho^{(f)}(\boldsymbol{X})\cdot\ddot{\boldsymbol{X}},
\end{align}
where the operators act to the right. In the first term of $\rho^{(2)}$, the leftmost $\partial_{\boldsymbol{X}}$ also acts on the parametric dependence of the $\mathcal{L}_t^{-1}$ to its right. The second-order correction thus contains a term quadratic in the driving velocity and a term linear in its acceleration.

\subsection{Heat and work in the slow-driving limit}

The power injected by the driving and the heat current flowing from the system into reservoir $\alpha$ are given by (see Ec. (9) in the main text)
\begin{equation}\label{heat-work-SM}
    P(t) = \mathrm{Tr}\Big\{\rho(t)\,\frac{d\mathcal{H}_{\mathrm{S}}(t)}{dt}\Big\},
    \qquad
    J_\alpha(t) = -\mathrm{Tr}\Big\{\mathcal{D}_\alpha[\rho(t)]\,\mathcal{H}_{\mathrm{S}}(t)\Big\}.
\end{equation}
We adopt the sign convention used in the main text: $J_\alpha = d\langle\mathcal{H}_\alpha\rangle/dt>0$ denotes energy transferred from the system to reservoir $\alpha$.
We now substitute the slow-driving expansion, Eq.~\eqref{slow-exp}, into these expressions and keep terms up to second order. Throughout, the superscript $(n)$ denotes the order in $1/\mathcal{T}$, and summation over repeated indices is implied.

For the power we have $P(t) = P^{(1)}(t) + P^{(2)}(t) + \ldots$, where
\begin{equation}
    P^{(1)}(t) = \mathrm{Tr}\Big\{\rho^{(f)}(\boldsymbol{X})\,\frac{d\mathcal{H}_{\mathrm{S}}(t)}{dt}\Big\}
    = \mathrm{Tr}\Big\{\rho^{(f)}(\boldsymbol{X})\,\partial_{\boldsymbol{X}}\mathcal{H}_{\mathrm{S}}(\boldsymbol{X})\Big\}\cdot\dot{\boldsymbol{X}}.
\end{equation}
Since both reservoirs are at the same temperature $T$, the frozen state is the Gibbs state and $\mathrm{Tr}\{\rho^{(f)}\,\partial_{\boldsymbol{X}}\mathcal{H}_{\mathrm{S}}\} = \partial_{\boldsymbol{X}}F_{\mathrm{eq}}(\boldsymbol{X})$, where $F_{\mathrm{eq}}$ is the equilibrium free energy. 
The integral of $P^{(1)}$ over a closed cycle then vanishes, since it corresponds to the work done by a conservative force. The leading dissipative contribution is
\begin{equation}
    P^{(2)}(t) = \mathrm{Tr}\Big\{\rho^{(1)}(t)\,\frac{d\mathcal{H}_{\mathrm{S}}(t)}{dt}\Big\}
    = \dot{\boldsymbol{X}} \cdot \underline{\boldsymbol{\Lambda}}(\boldsymbol{X}) \cdot \dot{\boldsymbol{X}},
\end{equation}
which is even under reversal of the driving protocol, $\dot{\boldsymbol{X}} \to -\dot{\boldsymbol{X}}$, and is the lowest-order term contributing to entropy production. Here we have introduced the matrix $\underline{\boldsymbol{\Lambda}}$ with elements
\begin{equation}\label{lambda_mat}
    \Lambda_{jl}(\boldsymbol{X}) = \mathrm{Tr}\Big\{\Big(\left[\mathcal{L}_t^{-1}\,\partial_{X_j}\right]\rho^{(f)}(\boldsymbol{X})\Big)\,\partial_{X_l}\mathcal{H}_{\mathrm{S}}(\boldsymbol{X})\Big\}.
\end{equation}
Integrating the power over one cycle and introducing the dimensionless time $\theta = t/\mathcal{T}$, with $\boldsymbol{X}' \equiv d\boldsymbol{X}/d\theta = \mathcal{T}\dot{\boldsymbol{X}}$, we obtain the dissipated work
\begin{equation}\label{w-bound}
    W^{(2)} = \int_0^{\mathcal{T}} \dot{\boldsymbol{X}} \cdot \underline{\boldsymbol{\Lambda}} \cdot \dot{\boldsymbol{X}}\,dt
    = \frac{1}{\mathcal{T}} \int_0^{1} \boldsymbol{X}' \cdot \underline{\boldsymbol{\Lambda}} \cdot \boldsymbol{X}'\,d\theta,
\end{equation}
which scales as $1/\mathcal{T}$ for a fixed cycle geometry.

Analogously, substituting Eq.~\eqref{slow-exp} into the heat current gives
\begin{equation}\label{expansionj}
    J_\alpha(t) = J^{(f)}_\alpha(t) + J^{(1)}_\alpha(t) + J^{(2)}_\alpha(t) + \ldots,
\end{equation}
where $J^{(f)}_\alpha = -\mathrm{Tr}\{\mathcal{D}_\alpha[\rho^{(f)}]\,\mathcal{H}_{\mathrm{S}}\}$ is the stationary current in the frozen state, which vanishes since both reservoirs are at the same temperature, and
\begin{align}\label{definitions}
    J^{(1)}_\alpha(t) &= -\mathrm{Tr}\Big\{\mathcal{D}_\alpha\big[\rho^{(1)}(t)\big]\,\mathcal{H}_{\mathrm{S}}(t)\Big\}
    = \boldsymbol{\Lambda}_\alpha(\boldsymbol{X}) \cdot \dot{\boldsymbol{X}}, \nonumber \\
    J^{(2)}_\alpha(t) &= -\mathrm{Tr}\Big\{\mathcal{D}_\alpha\big[\rho^{(2)}(t)\big]\,\mathcal{H}_{\mathrm{S}}(t)\Big\}
    = \dot{\boldsymbol{X}} \cdot \underline{\boldsymbol{\Omega}}^{(2)}_\alpha(\boldsymbol{X}) \cdot \dot{\boldsymbol{X}}
    + \boldsymbol{\Lambda}^{(2)}_\alpha(\boldsymbol{X}) \cdot \ddot{\boldsymbol{X}}.
\end{align}
Here, $\boldsymbol{\Lambda}_\alpha$ and $\boldsymbol{\Lambda}^{(2)}_\alpha$ are vectors with components
\begin{equation}\label{lambda-al}
    \Lambda_{\alpha,j}(\boldsymbol{X})
    = -\mathrm{Tr}\Big\{\mathcal{D}_\alpha\Big[\mathcal{L}_t^{-1}\partial_{X_j}\rho^{(f)}(\boldsymbol{X})\Big]\,\mathcal{H}_{\mathrm{S}}(\boldsymbol{X})\Big\},
    \qquad
    \Lambda^{(2)}_{\alpha,j}(\boldsymbol{X})
    = -\mathrm{Tr}\Big\{\mathcal{D}_\alpha\Big[\left[\mathcal{L}_t^{-1}\right]^2\partial_{X_j}\rho^{(f)}(\boldsymbol{X})\Big]\,\mathcal{H}_{\mathrm{S}}(\boldsymbol{X})\Big\},
\end{equation}
and $\underline{\boldsymbol{\Omega}}^{(2)}_\alpha$ is a matrix with elements
\begin{equation}\label{omega2}
    \Omega^{(2)}_{\alpha,jl}(\boldsymbol{X}) = -\mathrm{Tr}\Big\{\mathcal{D}_\alpha\Big[\left[\mathcal{L}_t^{-1}\,\partial_{X_j}\right]\left[\mathcal{L}_t^{-1}\,\partial_{X_l}\right]\rho^{(f)}(\boldsymbol{X})\Big]\,\mathcal{H}_{\mathrm{S}}(\boldsymbol{X})\Big\}.
\end{equation}

Integrating $J^{(1)}_\alpha$ over one cycle yields the geometric heat
\begin{equation}\label{Q1}
    Q^{\rm geo}_\alpha = \int_0^{\mathcal{T}} \boldsymbol{\Lambda}_\alpha \cdot \dot{\boldsymbol{X}}\,dt
    = \oint_C \boldsymbol{\Lambda}_\alpha(\boldsymbol{X}) \cdot d\boldsymbol{X} = \iint \nabla \times \boldsymbol{\Lambda}_\alpha(\boldsymbol{X}) \; d^2 X = \iint \mathcal{F}_\alpha(\boldsymbol{X}) \; d^2 X,
\end{equation}
which is independent of $\mathcal{T}$ and depends only on the closed curve $C$ described by $\boldsymbol{X}$ in parameter space, i.e., it is purely geometric.

At second order, the term involving the acceleration can be integrated by parts. Since $\boldsymbol{X}(t)$ is periodic, the boundary terms vanish and
\begin{equation}\label{hip_acc}
    \int_0^{\mathcal{T}} \Lambda^{(2)}_{\alpha,j}\,\ddot{X}_j\,dt
    = -\int_0^{\mathcal{T}} \frac{\partial \Lambda^{(2)}_{\alpha,j}}{\partial X_i}\,\dot{X}_i\,\dot{X}_j\,dt,
\end{equation}
which allows us to define
\begin{equation}
    \tilde{\Omega}^{(2)}_{\alpha,ij} = -\frac{\partial \Lambda^{(2)}_{\alpha,j}}{\partial X_i}.
\end{equation}
Hence, the total second-order dissipative kernel is
\begin{equation}\label{omega_completo}
    \underline{\boldsymbol{\Omega}}_\alpha = \underline{\boldsymbol{\Omega}}^{(2)}_\alpha + \underline{\tilde{\boldsymbol{\Omega}}}^{(2)}_\alpha,
\end{equation}
and the second-order heat exchanged with reservoir $\alpha$ per cycle reads
\begin{equation}\label{Q2}
    \frac{\mathcal{C}_\alpha}{\mathcal{T}}
    = \int_0^{\mathcal{T}} \dot{\boldsymbol{X}} \cdot \underline{\boldsymbol{\Omega}}_\alpha \cdot \dot{\boldsymbol{X}}\,dt,
    \qquad
    \mathcal{C}_\alpha = \int_0^{1} \boldsymbol{X}' \cdot \underline{\boldsymbol{\Omega}}_\alpha \cdot \boldsymbol{X}'\,d\theta,
\end{equation}
where $\mathcal{C}_\alpha$ is independent of $\mathcal{T}$ for a fixed cycle geometry. Note that only the symmetric parts of $\underline{\boldsymbol{\Lambda}}$ and $\underline{\boldsymbol{\Omega}}_\alpha$ contribute to the quadratic forms in Eqs.~\eqref{w-bound} and \eqref{Q2}.

\subsection{Thermodynamic consistency in the slow-driving limit}
The quantities introduced above satisfy the first law of thermodynamics, as discussed in Ref.~\cite{Caselli2026May}. The instantaneous internal energy is defined as
\begin{equation}\label{internal-energy}
    E_{\mathrm{S}}(t)
    = \mathrm{Tr}\!\left\{\rho(t)\mathcal{H}_{\mathrm{S}}(t)\right\},
\end{equation}
the first law reads
\begin{equation}\label{first-law-SM}
    \frac{dE_{\mathrm{S}}}{dt}
    = P(t)-\sum_\alpha J_\alpha(t).
\end{equation}
The minus sign follows from the convention adopted in Eq.~\eqref{heat-work-SM}, according to which $J_\alpha>0$ denotes heat transferred from the system to reservoir $\alpha$.

Substituting the slow-driving expansion of the density matrix, Eq.~\eqref{slow-exp}, into Eq.~\eqref{internal-energy}, we write
\begin{equation}\label{energy-expansion}
    E_{\mathrm{S}}(t)
    = E_{\mathrm{S}}^{(f)}(t)+E_{\mathrm{S}}^{(1)}(t)+\ldots,
    \qquad
    E_{\mathrm{S}}^{(n)}(t)
    = \mathrm{Tr}\!\left\{\rho^{(n)}(t)\mathcal{H}_{\mathrm{S}}(t)\right\},
\end{equation}
where $E_{\mathrm{S}}^{(f)}$ corresponds to $\rho^{(f)}$. Collecting the first-order terms in Eq.~\eqref{first-law-SM} gives
\begin{equation}\label{balance-order-1}
    \frac{dE_{\mathrm{S}}^{(f)}}{dt}
    = P^{(1)}(t)-\sum_\alpha J_\alpha^{(1)}(t).
\end{equation}
Integration over one driving period yields
\begin{equation}\label{integrated-balance-order-1}
    0
    = \int_0^{\mathcal{T}} P^{(1)}(t)\,dt
    -\sum_\alpha Q_\alpha^{\rm geo}.
\end{equation}
Since both reservoirs are at the same temperature, the first term vanishes because $P^{(1)}$ is generated by a conservative force. Therefore,
\begin{equation}\label{geometric-heat-conservation}
    \sum_\alpha Q_\alpha^{\rm geo}=0 , \qquad Q_L^{\rm geo} = -Q_R^{\rm geo} .
\end{equation}

At second order, the energy balance becomes
\begin{equation}\label{balance-order-2-local}
    \frac{dE_{\mathrm{S}}^{(1)}}{dt}
    = P^{(2)}(t)-\sum_\alpha J_\alpha^{(2)}(t).
\end{equation}
Using the expressions derived above and integrating over one period, we obtain
\begin{align}\label{balance-order-2}
    0
    &= \int_0^{\mathcal{T}}\frac{dE_{\mathrm{S}}^{(1)}}{dt}\,dt \nonumber\\
    &= \int_0^{\mathcal{T}}
    \dot{\boldsymbol{X}}\cdot\underline{\boldsymbol{\Lambda}}\cdot\dot{\boldsymbol{X}}\,dt
    -\sum_\alpha\int_0^{\mathcal{T}}
    \dot{\boldsymbol{X}}\cdot\underline{\boldsymbol{\Omega}}_\alpha
    \cdot\dot{\boldsymbol{X}}\,
    dt.
\end{align}
After integrating the acceleration term by parts, as in Eq.~\eqref{hip_acc}, this relation reduces to
\begin{equation}\label{integrated-energy-balance}
    W^{(2)}
    = \sum_\alpha\frac{\mathcal{C}_\alpha}{\mathcal{T}}.
\end{equation}
Thus, the total second-order heat delivered to the reservoirs equals the dissipated work performed by the drive.

\section{Full time-dependent Lindblad evolution}\label{full_lindblad}

Following Ref.~\cite{Cavina2017Aug}, we describe the full Lindblad evolution used to benchmark the collisional-model results shown in Fig. 2 (b) of the main text.

\subsection{Evolution operator}

We define the evolution operator $\mathcal{V}(t,t_0)$ through $\rho(t) = \mathcal{V}(t,t_0)\,\rho(t_0)$. Since $\dot{\rho} = \mathcal{L}_t\rho$ holds for any initial state, where $\mathcal{L}_t$ is the Liouvillian of Eq.~\eqref{lindblad}, $\mathcal{V}$ satisfies
\begin{equation}\label{eq_dif}
    \frac{d \mathcal{V}(t,t_0)}{dt} = \mathcal{L}_t\, \mathcal{V}(t,t_0), \qquad \mathcal{V}(t_0,t_0) = \mathbb{I}.
\end{equation}
Integrating, Eq.~\eqref{eq_dif} can be recast as the integral equation
\begin{equation}
   \mathcal{V}(t,t_0) = \mathbb{I} + \int_{t_0}^t dt^\prime\, \mathcal{L}_{t^\prime}\, \mathcal{V}(t^\prime,t_0),
\end{equation}
which, iterated starting from $\mathcal{V} = \mathbb{I}$, yields the Dyson series
\begin{equation}\label{dyson}
    \mathcal{V}(t,t_0) = \sum_{n=0}^\infty \frac{1}{n!} \int_{t_0}^t dt_1 \dots \int_{t_0}^t dt_n\, T_t[\mathcal{L}_{t_1}\dots \mathcal{L}_{t_n}] = T_t\, e^{\int_{t_0}^t dt^\prime \mathcal{L}_{t^\prime}} .
\end{equation}

\subsection{Piecewise evolution}

If $\mathcal{L}_t \equiv \mathcal{L}$ does not depend on time, the time ordering can be dropped and the exponent can be integrated directly, $\mathcal{V}(t,t_0) = e^{(t-t_0) \mathcal{L}}$.

Otherwise, if $\mathcal{L}_t$ depends on time, we partition the evolution into $N$ intervals $(t_{i-1},t_i)$, $i = 1,\dots,N$, of length $\Delta t = (t-t_0)/N$, with $t_i = t_0 + i\,\Delta t$. From the definition of $\mathcal{V}$ it follows that $\mathcal{V}(t,t_0) = \mathcal{V}(t,s)\,\mathcal{V}(s,t_0)$ for any $t_0 \leq s \leq t$, which allows us to write
\begin{equation}\label{composition}
    \mathcal{V}(t_N,t_0) = \mathcal{V}(t_N,t_{N-1}) \cdots \mathcal{V}(t_2,t_1)\, \mathcal{V}(t_1,t_0) .
\end{equation}
Within each interval we approximate $\mathcal{L}_t$ by its value at the beginning of the interval, $\mathcal{L}_t \simeq \mathcal{L}_{t_{i-1}}$. The time ordering in Eq.~\eqref{dyson} then becomes trivial and the series sums to
\begin{equation}\label{one-step}
    \mathcal{V}(t_i,t_{i-1}) \simeq e^{\Delta t\, \mathcal{L}_{t_{i-1}}} .
\end{equation}
This approximation introduces an error of order $\Delta t^2$ in $\mathcal{V}(t_i,t_{i-1})$, dominated by the variation of $\mathcal{L}_t$ within the interval; the time-ordering corrections involve commutators $[\mathcal{L}_{t_1},\mathcal{L}_{t_2}]$ and are of higher order. Since there are $N = (t-t_0)/\Delta t$ intervals, the accumulated error is $\mathcal{O}(\Delta t)$ and vanishes for $\Delta t \to 0$.

Substituting Eq.~\eqref{one-step} into Eq.~\eqref{composition}, the evolution of the density matrix can be computed step by step, starting from the initial state $\rho(t_0)$, as
\begin{equation}
    \rho(t_i) = e^{\Delta t\, \mathcal{L}_{t_{i-1}}}\, \rho(t_{i-1}) , \qquad i = 1,\dots,N ,
\end{equation}
which reproduces the exact evolution in the limit $\Delta t \to 0$.

Substituting the resulting $\rho(t)$ into Eq. (9) of the main text and integrating over one driving period yields the heat and work per cycle, which become exact in the limit $\Delta t \to 0$ and serve as benchmarks for the collision model.

\bibliography{bibliography.bib}